\documentclass[aps,prl,superscriptaddress,reprint]{revtex4-1}
\usepackage{amsmath}
\usepackage{amssymb}
\usepackage{graphicx}
\usepackage{hyperref}
\usepackage{url}
\usepackage{color,xcolor}
\usepackage{ulem}
\usepackage{float}
\usepackage{epstopdf}
\begin{document}

\title{Magnetic states of iron-based two-leg ladder tellurides}
\author{Yang Zhang}
\author{Ling-Fang Lin}
\affiliation{Department of Physics and Astronomy, University of Tennessee, Knoxville, TN 37996, USA}
\affiliation{School of Physics, Southeast University, Nanjing 211189, China}
\author{Adriana Moreo}
\affiliation{Department of Physics and Astronomy, University of Tennessee, Knoxville, TN 37996, USA}
\affiliation{Materials Science and Technology Division, Oak Ridge National Laboratory, Oak Ridge, TN 37831, USA}
\author{Shuai Dong}
\affiliation{School of Physics, Southeast University, Nanjing 211189, China}
\author{Elbio Dagotto}
\email{Corresponding author: edagotto@utk.edu}
\affiliation{Department of Physics and Astronomy, University of Tennessee, Knoxville, TN 37996, USA}
\affiliation{Materials Science and Technology Division, Oak Ridge National Laboratory, Oak Ridge, TN 37831, USA}

\date{\today}

\begin{abstract}
The recent discovery of superconductivity at high pressure
in the two-leg ladder compounds BaFe$_2X_3$ ($X$=S, Se)
started the novel field of quasi-one-dimensional iron-based superconductors. In this publication,
we use Density Functional Theory (DFT) to predict that the previously barely explored ladder
compound RbFe$_2$Te$_3$ should be magnetic with a CX-type arrangement involving
ferromagnetic rungs and antiferromagnetic legs, at the realistic density of $n=5.5$ electrons per iron.
The magnetic state similarity with BaFe$_2$S$_3$ suggests that RbFe$_2$Te$_3$ could also become
superconducting under pressure. Moreover, at $n=6.0$ our DFT phase diagrams (with and without
lattice tetramerization) reveal that the stable magnetic states could be either
a 2$\times$2 magnetic Block-type, as for $X$=Se, or a previously never observed  before CY-type state,
with ferromagnetic legs and antiferromagnetic rungs.
In the Te-based studies, electrons are more localized than in S, implying that the degree
of electronic correlation is enhanced for the Te case.
\end{abstract}

\maketitle

\section{Introduction}
Although the first high critical temperature iron-based superconductors were discovered more than a decade ago,
the origin of its pairing mechanism is still highly debated and the topic remains one of the most important open problems  in Condensed Matter Physics~\cite{Stewart:Rmp,Dagotto:Rmp,Dai:Rmp,Dai:Np}. It is widely believed that the crystal structure, magnetic properties, and the degree of electronic correlation are all fundamental aspects to clarify the physics of these materials~\cite{Mazin:np,Johnston:Ap,Dagotto:Rmp,Si:NRM}. For the vast majority of initially reported iron-based superconductors, the crystal structures consisted of slightly distorted two-dimensional (2D) iron square lattices made of Fe$X_4$ tetrahedra ($X$ = pnictides or chalcogens)
~\cite{Li:Np,Bao:Cpl,Johnston:Ap}. The electronic correlation effects can not be neglected~\cite{Dai:Np}, causing many novel physical features, such as Fermi surfaces without hole pockets, complex magnetic spin orders, as well as orbital selective Mott states~\cite{Dai:Np,Dagotto:Rmp}.

Recently, the discovery of superconductivity in the so-called $123$-type compounds BaFe$_2X_3$ ($X$=S/Se) opened a new branch of research in iron-based superconductors~\cite{Takahashi:Nm,Ying:prb17}. Different from the 2D iron square lattice arrangement, the $123$-type iron  chalcogenides display a dominant quasi-one-dimensional two-leg ladder crystal structure that
has been much analyzed~\cite{Saparov:Prb,Lei:Prb,du:prb12,luo:prb13,Dong:PRL14,mourigal:prl15}. These recent developments resemble the discovery in the 90's of superconductivity in Cu-oxide ladders~\cite{cu-ladder1,cu-ladder2,cu-ladder3} that also opened a fertile area of research.

Under ambient conditions, BaFe$_2$S$_3$ displays CX stripe antiferromagnetic (AFM) order -- AFM along the legs and ferromagnetic (FM) along the rungs -- below $120$~K with a magnetic moment $\sim 1.2$ $\mu_{\rm B}$/Fe~\cite{Takahashi:Nm,chi:prl}. This magnetic moment is smaller than the theoretical maximum value of $4$ $\mu_{\rm B}$/Fe,
obtained by considering the high-spin $S = 2$ configuration and the electronic density $n = 6.0$
for BaFe$_2$S$_3$. Superconductivity was observed at $P\sim 11$~GPa with the highest  critical temperature $T_{\rm c}$ being $24$~K ~\cite{Takahashi:Nm,Yamauchi:prl15}. Since then, several experimental and theoretical studies have followed~\cite{Wang:prb16,Patel:prb17,Pizarro:prm,Zheng:prb18,König:prb18}. Above $10$~GPa, a metal-insulator transition (MIT) and associated first-order magnetic phase transition were recently observed for BaFe$_2$S$_3$~\cite{Zhang:prb17,Materne:prb19}. One possible explanation is that pressure changes the bandwidth of these materials, thus
altering the degree of correlation~\cite{Takahashi:Nm,Ying:prb17}. An alternative,
based on model calculations, is that high pressure could change the Fe electronic density, effectively doping the two-leg ladders. In fact, calculations based on the density matrix renormalization group~\cite{Patel:prb16} observed clear tendencies to form Cooper pairs at intermediate
Hubbard coupling strengths upon light doping. Similar self-doping effects under pressure were also obtained using DFT calculations~\cite{Zhang:prb17}.

BaFe$_2$Se$_3$ is another recently discovered superconducting ladder under high pressure~\cite{Ying:prb17}. Without external pressure, BaFe$_2$Se$_3$ is an AFM Mott insulator and displays an exotic Block-type magnetic
order below $\sim 256$ K, with a robust local magnetic moment $\sim 2.8$ $\mu_{\rm B}$/Fe~\cite{Saparov:Prb,Caron:Prb12,Lei:Prb,Nambu:Prb}. This material is theoretically predicted to be multiferroic~\cite{Dong:PRL14} and recently confirmed to be polar at high temperature~\cite{Aoyama:prb19}. In particular, BaFe$_2$Se$_3$ is in an orbital-selective Mott phase (OSMP) according to neutron experiments at ambient pressure~\cite{mourigal:prl15}.
Moreover, there are several other two-leg ladder iron chalcogenides, with almost all the studies focusing on iron sulfides and selenides. For example, KFe$_2$Se$_3$ was observed to have a CX-type stripe AFM order~\cite{Caron:Prb12}, similarly as CsFe$_2$Se$_3$~\cite{chi:prl} and RbFe$_2$Se$_3$~\cite{Wang:prb16}. In particular, the KFe$_2$S$_3$ compound was predicted to display a first-order transition under high pressure in our recent work~\cite{Zhang:prb17}.

\begin{figure}
\centering
\includegraphics[width=0.48\textwidth]{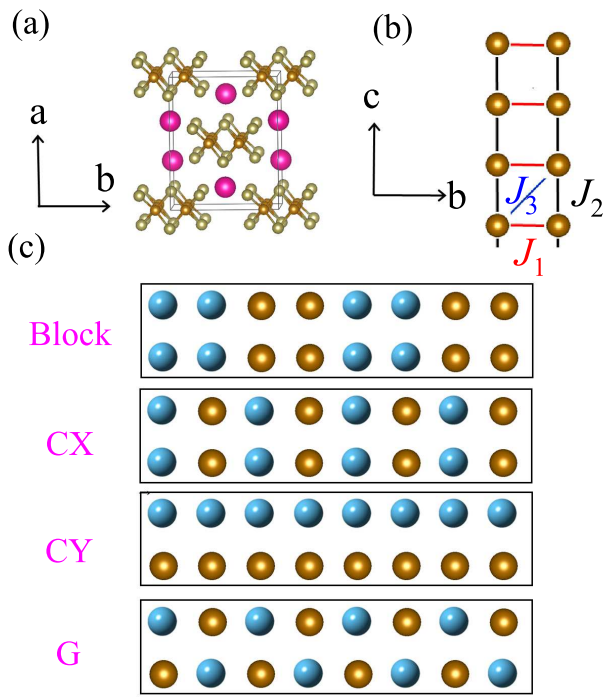}
\caption{(a) Schematic crystal structure of RbFe$_2$Te$_3$ (electronic density $n = 5.5$) with the convention: Pink = Rb; Brown = Fe; Dark Yellow = Te. (b) Sketch of one $Cmcm$ ladder. The iron-iron distance is uniform along the legs.
(c) Sketch of possible antiferromagnetic magnetic orders in each individual ladder studied here. Spin up and spin down are distinguished by different colored balls.}
\label{Fig1}
\end{figure}

Considering the columns of pnictogens and chalcogens in the periodic table, the next natural step in iron ladders is to move one row down and use Sb or Te. However, surprisingly there is virtually no experimental literature available using these elements. For Te, we are only aware of $one$ publication many years ago where it was reported that RbFe$_2$Te$_3$ also favors the $Cmcm$ crystal structure~\cite{Klepp:JOAC}, similar to BaFe$_2$S$_3$, where the iron-iron distances are uniform in the non-tilting ladder, as shown in Fig.~\ref{Fig1}. In RbFe$_2$Te$_3$, the valence of Fe is $+2.5$ indicating that the realistic density is $n=5.5$ electrons per iron considering the 4s$^2$3d$^6$ configuration in the Fe atoms. It is important to remark that there are still no $n=5.5$ ladders being reported to be superconducting under high pressure.


In the present publication, the magnetic properties and electronic structure corresponding to RbFe$_2$Te$_3$ are studied based on first-principles DFT calculations. The CX-type spin order is predicted to be the most likely magnetic ground state in our $n=5.5$ DFT phase diagrams. For comparison, for the $n=6.0$ BaFe$_2$Se$_3$ compound
the 2$\times$2 magnetic Block-type state was found to be stable after including lattice tetramerization. In the Te-based compound, we found that electrons are more localized than in S, implying that the degree of electronic correlation is enhanced for the Te case. Future experimental efforts should be devoted to this interesting Te-ladder compound.

\section{Method}
The first-principles DFT calculations used here were performed with the projector augmented-wave (PAW) potentials as implemented in the Vienna {\it ab initio} Simulation Package (VASP) code~\cite{Kresse:Prb,Kresse:Prb96}. The Perdew-Burke-Ernzerhof (PBE) exchange function was employed~\cite{Perdew:Prl} and the plane-wave cutoff energy was $500$ eV. Since different magnetic configurations have different minimal unit cells, the mesh was appropriately modified for all the candidates to render the $k$-point densities approximately the same in reciprocal space, i.e. $6\times6\times8$ for Block-type and $6\times5\times10$ for FM-type. In addition, we have tested that these $k$-point meshes already lead to converged energies when compared with denser meshes.

As a first step, we considered the spin polarized version of the generalized gradient approximation (GGA) potential~\cite{Perdew:Prl} to study the lattice ground-state properties of bulk RbFe$_2$Te$_3$. Since the PBE-GGA function is known to give an accurate description of the two-leg ladder systems ~\cite{Zhang:prb17,Zhang:prb18,Suzuki:prb}, we do not consider the effective Hubbard $U_{\rm eff}$. Both the lattice constants and atomic positions were fully relaxed with different spin configurations until the force on each atom was below $0.005$ eV/{\AA}.

To understand magnetism, we adopted the Local Density Approximation (LDA) + $U$ method ~\cite{Liechtenstein:prb}, where the on-site Coulomb interaction $U$ and on-site exchange interaction $J$ were considered. To alleviate the computing time required, we just considered the $(0, 0, 0)$ order between ladders with a minimum unit cell to obtain the phase diagram of the $n=5.5$ and $n=6.0$ ladders. Due to the dominance of the in-ladder magnetic order, the magnetic correlations between ladders can only slightly affect the energies and physical properties.

The generalized LDA+$U$ functional is the following~\cite{Liechtenstein:prb,Anisimov:JPCM}:
\begin{eqnarray}
\nonumber E^{\rm LDA+U}[\rho^\sigma(r),[\{n^\sigma\}]&=&E^{\rm LSDA}[\rho^\sigma(r)]+E^U[\{n^\sigma\}]\\
&&-E_{\rm dc}[\{n^\sigma\}],
\end{eqnarray}
where $\rho^\sigma(r)$ is the charge density for electrons with spin projection $\sigma$,
while $\{ n^\sigma \}$ are the elements of the density matrix. Here, the density matrix is defined as:
\begin{equation}
n^{\sigma}_{mm'}=-1/\pi \int^{\rm E_F}ImG^{\sigma}_{ilm,ilm'}(E)dE,
\end{equation}
where $i$ denotes site, $l$ indicates the orbital quantum number, and $m$ the spin number. Note that there is summation for $i$ and $l$ implicit, $G^{\sigma}_{ ilm,ilm'}(E)=<ilm\sigma|(E-H)^{-1}|ilm'\sigma>$ are the matrix elements of the Green function matrix in the localized representation, and $H$ is the effective single-electron Hamiltonian. The orbital polarizations are absent in the  LSDA first term, and the second term in Eq.(1) can be described by the Hartree-Fock (HF) mean-field theory~\cite{Liechtenstein:prb,Anisimov:JPCM}:
\begin{eqnarray}
\nonumber E^U[\{n\}]&=&1/2\sum_{\{m\},\sigma}\{<m,m''|V_{\rm ee}|m',m'''>n^{\sigma}_{mm'}n^{-\sigma}_{m''m'''}\\
\nonumber &&+(<m,m''|V_{\rm ee}|m',m'''>-\\
&&<m,m''|V_{\rm ee}|m''',m'>)n^{\sigma}_{mm'}n^{\sigma}_{m''m'''}\},
\label{1}
\end{eqnarray}
where $V_{\rm ee}$ are the screened Coulomb interactions among $nl$-electrons. The double counting term ($E_{\rm dc}$) is described by
\begin{equation}
E_{\rm dc}[\{n^{\sigma}\}]=1/2Un(n-1)-1/2J[n^{\uparrow}(n^{\uparrow}-1)+n^{\downarrow}(n^{\downarrow}-1)],
\end{equation}
where $n^\sigma = Tr(n^{\sigma}_{mm'})$ and $n= n^{\uparrow}+n^{\downarrow}$. $U$ and $J$ are the Coulomb interaction and exchange interaction, respectively. If the density matrix becomes diagonal, the present rotationally-invariant method is equivalent to the ordinary LDA+$U$ approach ~\cite{Anisimov:prb}.

\section{Results}

\subsection{Physical properties of RbFe$_2$Te$_3$}
To find out what magnetic configuration becomes the ground state of RbFe$_2$Te$_3$, we adopted the spin polarized method within the GGA potential to fully relax the crystal lattices and atomic position since the PBE-GGA function was widely used in previous DFT calculations of two-leg iron ladder systems~\cite{Zhang:prb17,Zhang:prb18,Suzuki:prb,Zheng:prb18}. Various possible (in-ladder) magnetic arrangements were imposed on the iron ladders [see Fig.~\ref{Fig1}(c)], such as non-magnetic (NM), FM, AFM with FM rungs and AFM legs (CX), AFM with AFM rungs and FM legs (CY),
AFM in both rung and leg directions (G), and 2$\times$2 Block-AFM (Block)~\cite{Zhang:prb17}. Furthermore,
the ($\pi$, $\pi$, $0$) order between ladders was adopted, as suggested by neutron scattering results~\cite{Caron:Prb12} for K$_x$Ba$_{\rm 1-x}$Fe$_2$Se$_3$. Our main results for RbFe$_2$Te$_3$ are summarized in Table~\ref{Table1}.

Under ambient conditions, our DFT calculations performed for several magnetic candidates [the tested spin configurations are shown in Fig.~\ref{Fig1}(c)] indicate that the CX-type magnetic order is the most stable ground state of the ensemble used. For this CX-type state, the calculated local magnetic moment of Fe is about $2.71$ $\mu_{\rm B}$/Fe. It should be noted that it is quite common to overestimate the local magnetic moment when using the spin polarized method within the GGA potential in calculations of iron-based superconductors~\cite{Mazin:np,Suzuki:prb,Zhang:prb17}, which could be caused by the coexistence of localized Fe spins and itinerant electrons~\cite{Ootsuki:prb}. Another possibility is the existence of strong quantum zero-point fluctuations in this quasi-one-dimensional two-leg ladder system. For comparison, the calculated local magnetic moment for BaFe$_2$S$_3$ and KFe$_2$Se$_3$ are $2.08$ $\mu_{\rm B}$/Fe~\cite{Suzuki:prb,Zhang:prb17} and $2.65$ $\mu_{\rm B}$/Fe~\cite{Zhang:prb17}, respectively, which are larger than the experimental values $1.2$ $\mu_{\rm B}$/Fe and $2.1$ $\mu_{\rm B}$/Fe~\cite{Takahashi:Nm,Caron:Prb}. Hence, it is reasonable to assume that the experimental magnetic moment would be smaller than our calculated value
for RbFe$_2$Te$_3$.

The DFT calculated energy gap corresponding to the CX-type AFM order is about $0.39$ eV, which is close to the activation gap reported for CsFe$_2$Se$_3$~\cite{du:prb12}. This calculated gap for RbFe$_2$Te$_3$ is larger than the experimental value of BaFe$_2$S$_3$  $\sim 0.06-0.07$ eV ~\cite{Gonen:Cm}. According to the empirical knowledge gathered on iron-ladders, the larger gap indicates that a much higher pressure will be needed in the Te case
to achieve an insulator-metal transition, or to suppress magnetism, than in the S or Se cases.

\begin{table}
\centering\caption{The optimized
lattice constants ({\AA}), local magnetic moments (in $\mu_{\rm B}$/Fe units) within the default PAW sphere,
and band gaps (eV) for the various magnetic configurations,
as well as the energy differences (meV/Fe) with respect to the CX configuration taken as the
reference of energy. The experimental values (Exp. for short) are also listed for comparison.}
\begin{tabular*}{0.48\textwidth}{@{\extracolsep{\fill}}llllc}
\hline
\hline
  & $a$/$b$/$c$ & $M$ & Gap  & Energy \\
\hline
NM       & 12.665/9.953/5.683  & 0    & 0  & 396   \\
FM       & 13.164/10.625/5.629 & 2.64 & 0  & 238    \\
CX       & 12.803/10.233/5.868 & 2.71 & 0.39  & 0 \\
CY       & 12.622/10.522/5.653 & 2.43 & 0 & 236   \\
G        & 12.771/10.326/5.795 & 2.54 & 0 & 90    \\
Block    & 13.008/10.454/5.570 & 2.45 & 0 & 158   \\
Exp.     & 12.486/10.126/5.921 & --   & --&--\\
\hline
\hline
\end{tabular*}
\label{Table1}
\end{table}

Considering the intra-ladder magnetic order, the magnetism of RbFe$_2$Te$_3$ could be described
by a simple Heisenberg model:
\begin{equation}
H_{\rm spin}=-J_1\sum_{<i,j>}\textbf{S}_i\cdot\textbf{S}_j-J_2\sum_{[k,l]}\textbf{S}_k\cdot\textbf{S}_l-J_3\sum_{\{m,n\}}\textbf{S}_m\cdot\textbf{S}_n,
\end{equation}
where $J_1$ and $J_2$ are the exchange interactions in the rung and leg directions, respectively, while $J_3$ is the exchange coupling along the plaquette diagonal of iron atoms [Fig.~\ref{Fig1}(b)]. By fitting the DFT energies of various magnetic states, all the coefficients of this Heisenberg model can be obtained: $S^2J_1=44.2$ meV, $S^2J_2=-96.1$ meV, and $S^2J_3=-23.1$ meV, respectively ~\cite{Jcontext}. Similar to two-dimensional magnetic stripe iron superconductors and other two-leg iron ladders~\cite{Wang:prb16,Harriger:prb,Dai:Rmp}, they all display that the magnitude of the FM rung exchange coupling is smaller than the
magnitude of the AFM leg coupling.

\begin{figure}
\centering
\includegraphics[width=0.48\textwidth]{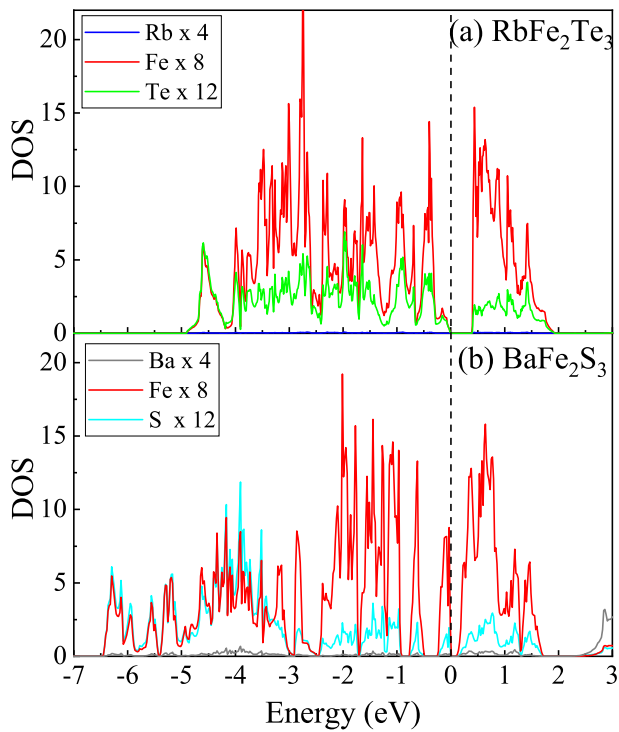}
\caption{DOS near the Fermi level using the CX-AFM states ($\pi$, $\pi$, $0$) for (a) RbFe$_2$Te$_3$ (at electronic density $n = 5.5$) and BaFe$_2$S$_3$ (at electronic density $n = 6.0$), respectively.
Blue= Rb; Black = Ba; Red = Fe; Green = Te; Cyan = S.}
\label{Fig2}
\end{figure}

According to the calculated density of states (DOS) of the CX-type AFM order of RbFe$_2$Te$_3$ [see Fig.~\ref{Fig2}(a)], the bands near the Fermi level are mainly contributed by Fe-$3d$ orbitals which are hybridized with Te-$5p$ orbitals. For comparison, we displayed the DOS of the CX-type AFM state of BaFe$_2$S$_3$ in Fig.~\ref{Fig2}(b). The bandwidth of the five iron bands of RbFe$_2$Te$_3$ ($\sim 6.8$ eV) is smaller than BaFe$_2$S$_3$ ($\sim 8$ eV), which indicates that effectively the iron orbitals
in RbFe$_2$Te$_3$ are more localized than in BaFe$_2$S$_3$. As remarked below, remember that the Fe-Fe
effective hopping is mediated by Te as a bridge, thus iron bandwidths are a consequence of Fe-Te-Fe hoppings.
It is interesting that the weight of Fe and Te near the Fermi level are smaller than in the case Fe and S. One possible reason is that RbFe$_2$Te$_3$ has $0.5$ electrons less than BaFe$_2$S$_3$ per iron ion, resulting in fewer iron states in RbFe$_2$Te$_3$.


\subsection{Magnetic phase diagrams for two-leg ladders at electronic densities $n = 5.5$ and $n=6.0$}

To understand better the magnetic properties of two-leg iron ladders, we used the LDA+$U$ method with GGA potential to compare different spin configurations by changing the on-site Coulomb interaction $U$ and on-site exchange interaction $J$. Here, to save computing resources, the (0, 0, 0) order between ladders was considered because the in-ladder magnetic coupling is dominant in two-leg iron systems.

\begin{figure}
\centering
\includegraphics[width=0.46\textwidth]{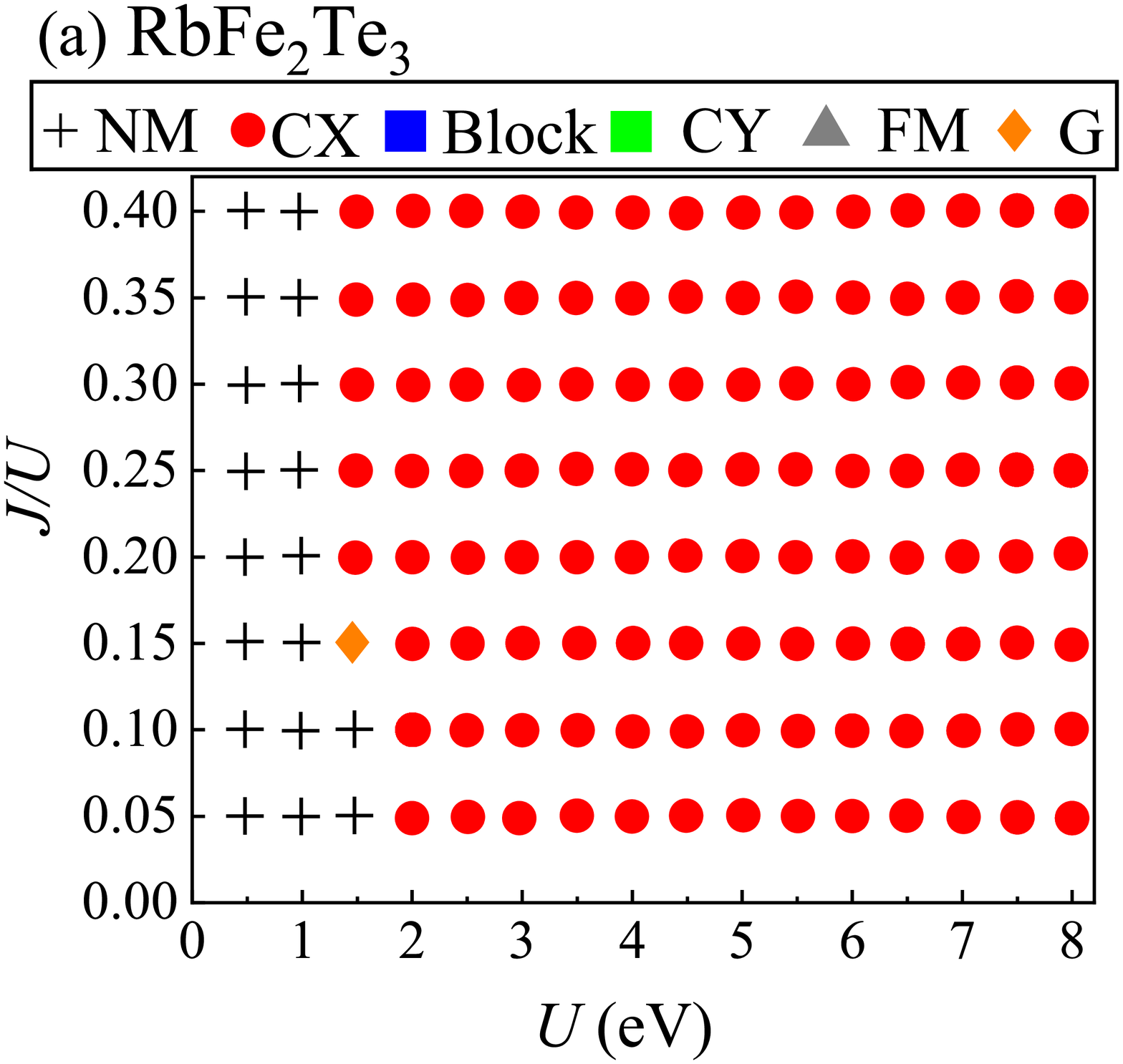}
\includegraphics[width=0.46\textwidth]{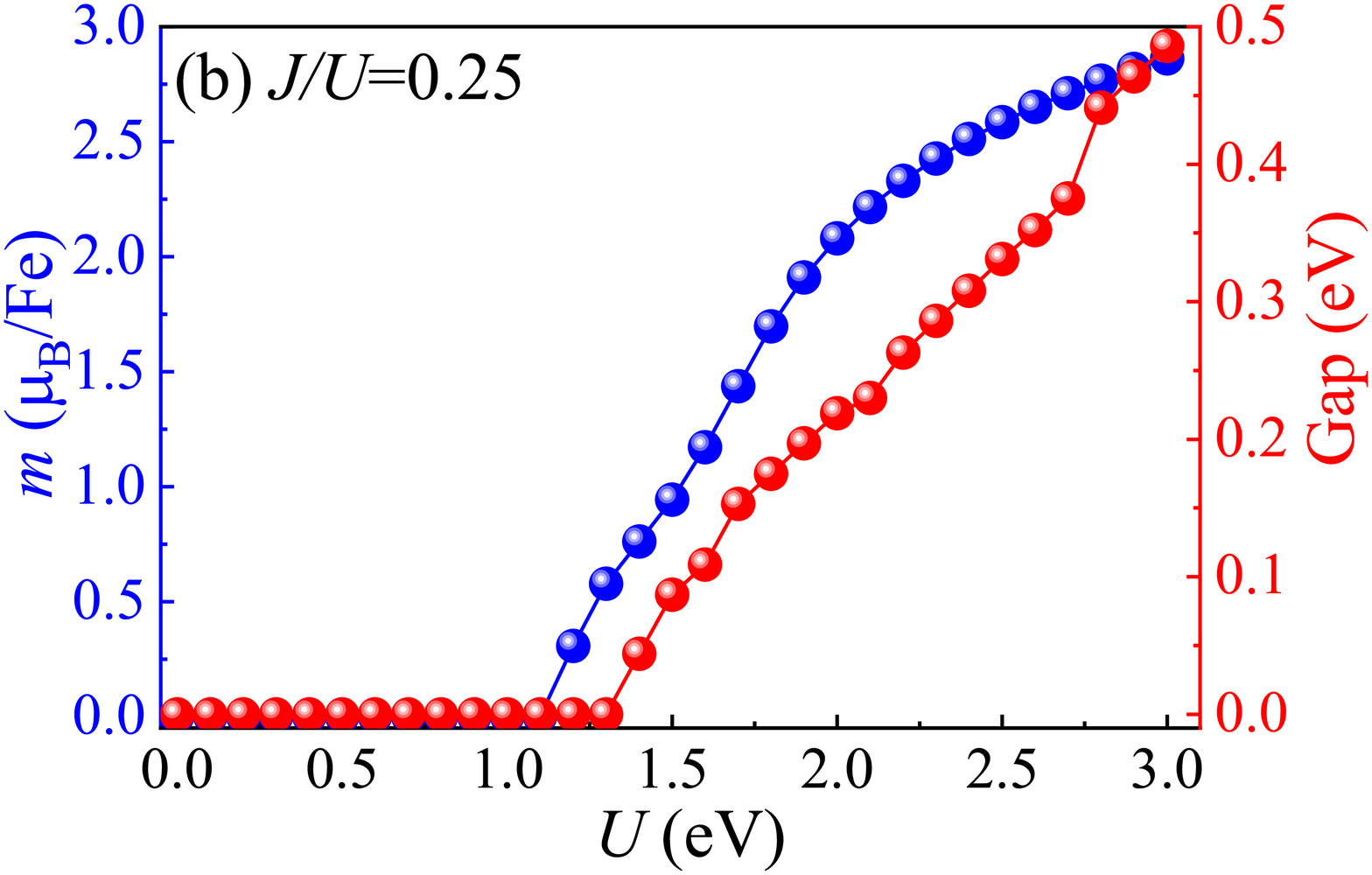}
\caption{(a) Phase diagram based on the experimental lattice constants of RbFe$_2$Te$_3$, employing the LDA+$U$ technique at the electronic density $n = 5.5$. (b) Evolution of local magnetic moments and band gaps of RbFe$_2$Te$_3$ for the CX-type AFM magnetic state, as a function of $U$, at $J/U = 0.25$.}
\label{Fig3}
\end{figure}

Let us start our description of the main results considering the ladder electronic density $n=5.5$,
corresponding to RbFe$_2$Te$_3$, using periodic boundary conditions, based on the experimental
crystal structure~\cite{Klepp:JOAC}. As shown in Fig.~\ref{Fig3}(a), there is only one magnetic state (CX-type) stable in our phase diagram [except for one anomalous point ($U=1.5$ eV and $J/U= 0.15$)] even when the Hubbard coupling $U$ and exchange interaction $J$ are varied in a wide range. This clearly indicates that the CX-type order is quite stable in our $n=5.5$ phase diagram, which is consistent with existing studies of magnetism in $n=5.5$ iron ladders~\cite{Caron:Prb12,Takahashi:Nm,Wang:prb16}. Hence, we arrive to the reasonable conclusion that CX-type AFM is the most likely magnetic ground state of
Te-based ladders, and likely other iron ladders with electronic density $n = 5.5$. This CX state of iron ladders can be considered quite similar to the prevalent stripe C-AFM order of iron 2D layered systems~\cite{Dai:Np,zhang:prm,Sasmal:prl}.

To qualitatively describe the Mott insulator of RbFe$_2$Te$_3$, we calculated the magnetic moment and energy gap by increasing $U$ at the realistic $J/U=0.25$~\cite{Dai:Np}, as displayed in Fig.~\ref{Fig3}(b). When $U$ is small, the magnetic moment of iron is zero, all the iron bands overlap, and the system is in a metallic state. By increasing
$U$ to a critical value, the spin up and down bands split, resulting in the CX-type AFM order while the system is still metallic in a very narrow $U$ range near 1.25 eV. Continuing to increase $U$, the valence band and the conduction band separate from each other opening a gap, and producing an insulating phase. Our results for $J/U=0.25$ qualitatively describe the Mott metal-insulator phase transition. In our LDA+$U$ approximation, the Hubbard $U$ splits the the spin up/down near $U = 1.2$ eV, and opens the gap at $U = 1.4$ eV.

\begin{figure}
\centering
\includegraphics[width=0.46\textwidth]{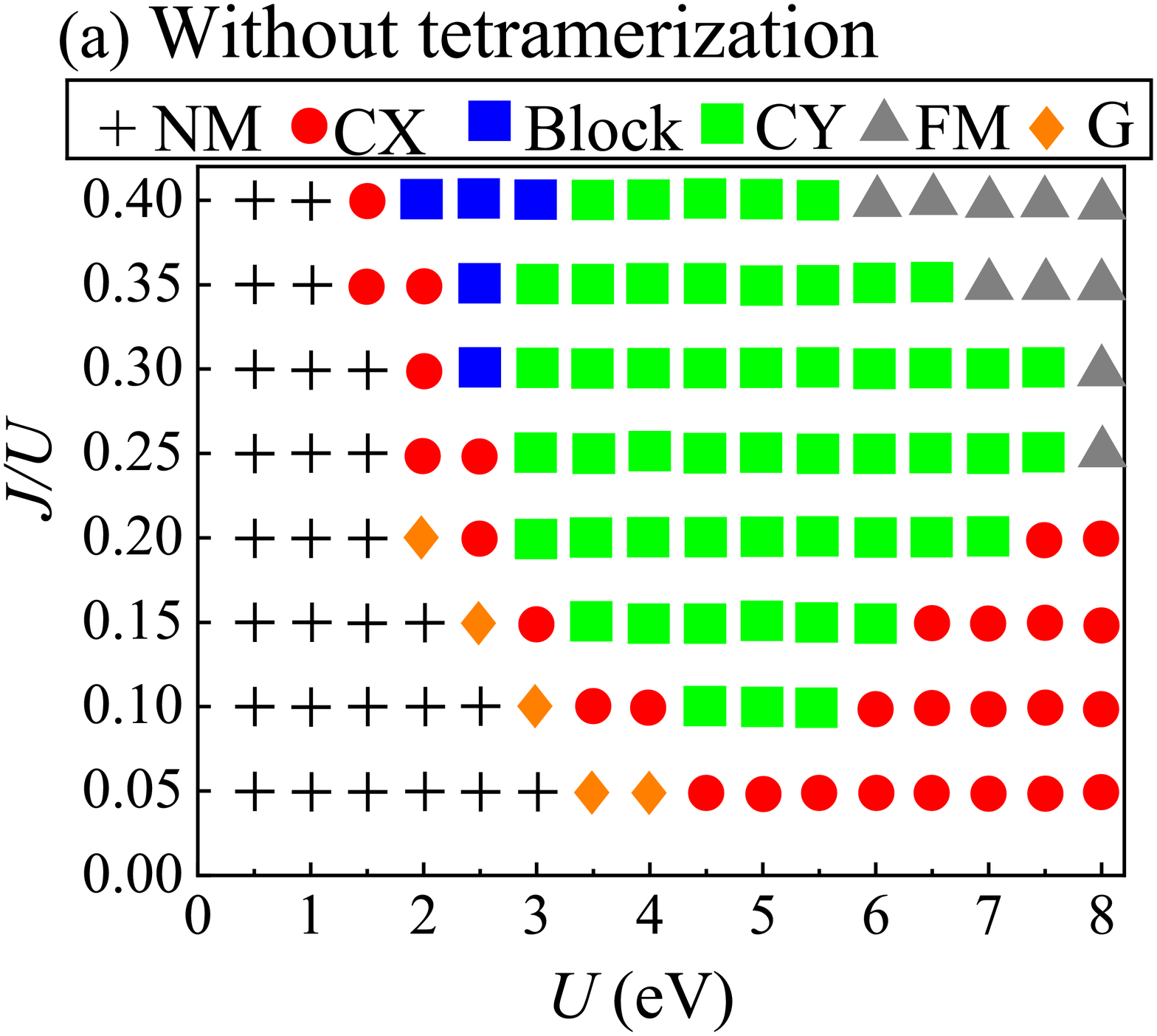}
\includegraphics[width=0.46\textwidth]{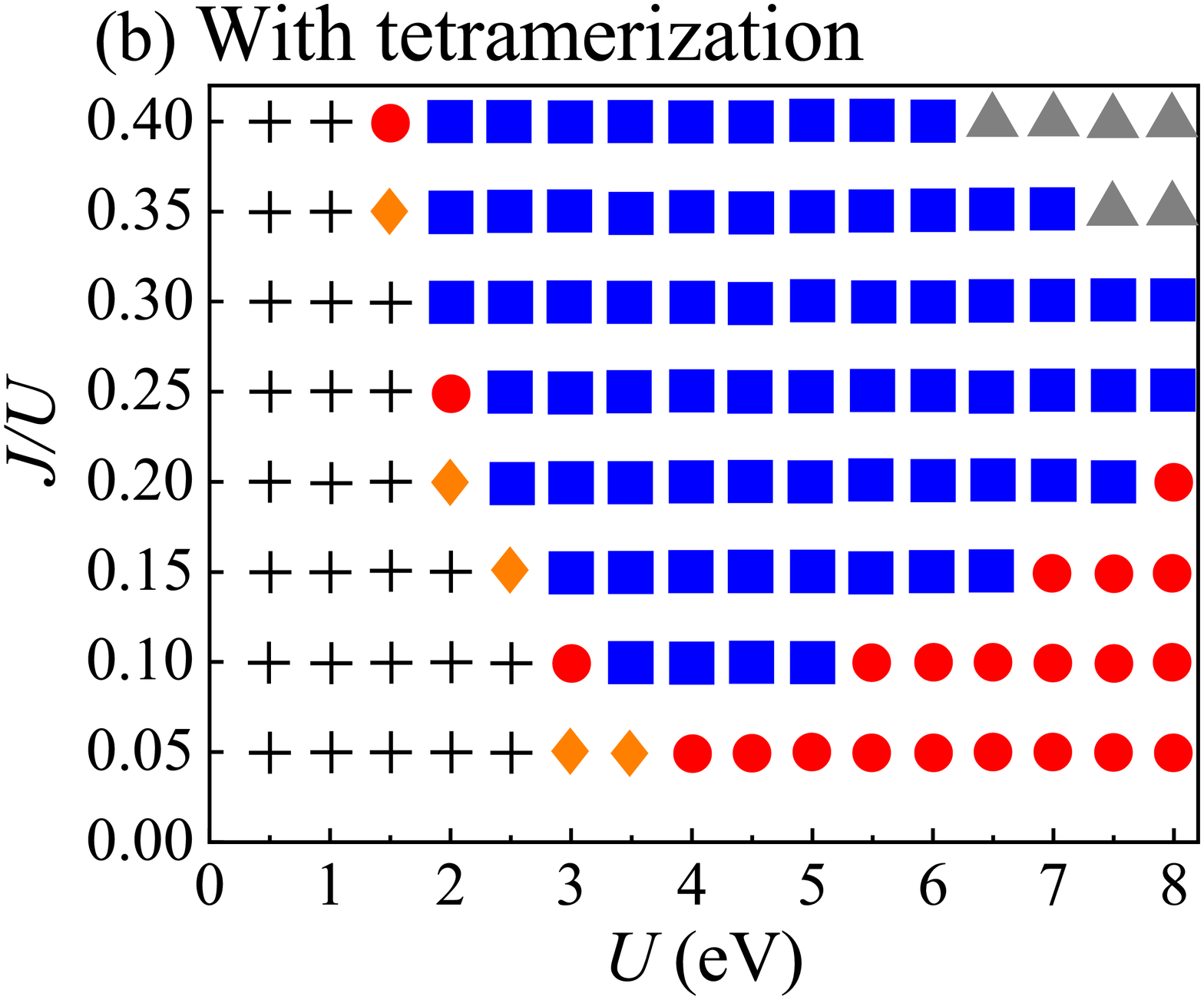}
\includegraphics[width=0.46\textwidth]{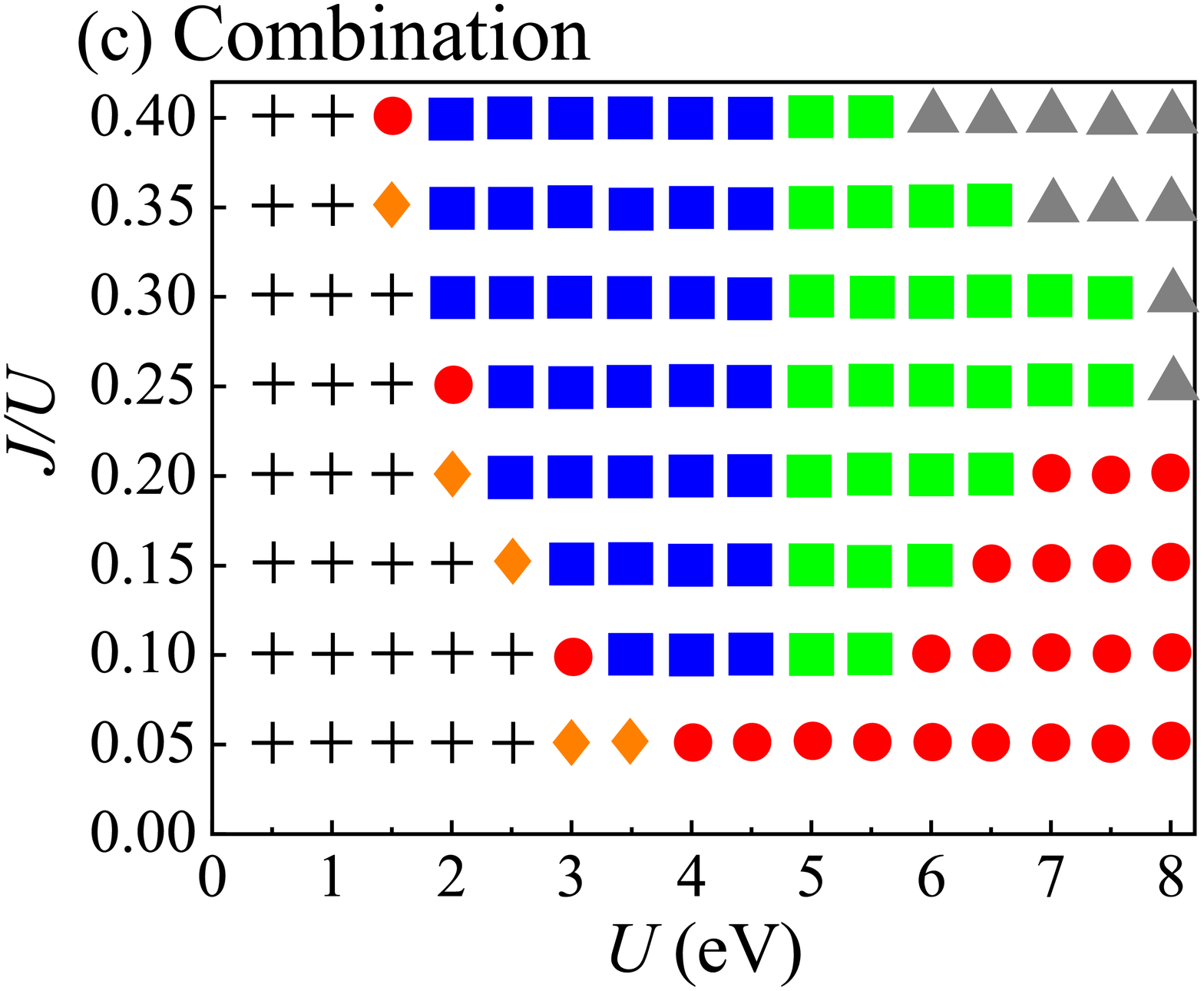}
\caption{(a-c) Phase diagrams of BaFe$_2$Se$_3$, employing the LDA+$U$ technique with electronic density $n = 6.0$. (a) Without tetramerization. (b) With tetramerization. (c) Combination, which is the most reliable prediction, obtained by comparing energies with and without tetramerization.}
\label{Fig4}
\end{figure}

Let us consider now the results for ladders with electronic
density $n=6.0$, corresponding to BaFe$_2$Se$_3$, which
experimentally is known to display the exotic 2$\times$2 Block-type AFM order~\cite{Caron:Prb12}.
As a first step, we use the crystal structure $without$ lattice tetramerization. Here, we adopted the crystal structure based on the $Pnma$ phase obtained from neutron experiments~\cite{Caron:Prb} which does not consider the magnetic exchange striction effect.
The phase diagram that we obtained for $n=6.0$ iron ladders become far richer than at $n=5.5$,
as displayed in Fig.~\ref{Fig4}(a), including five different magnetic states, with a surprising dominance
of the CY state, followed by CX with regards to area covered in the phase diagram~\cite{note-added}.
Note that here there is only a small region of the Block-type order in our DFT phase diagram, although
this state is the one found experimentally in Se-based ladders. However, it should remarked that the energy between Block-type and CY-type is less than $10$ meV/Fe
at $U = 3$~eV and $3.5$ eV. In other regions of our phase diagram, the energy of the Block-type remains
only slightly higher than the energy of the CY-type. To address better this issue note that
the Block-type AFM order naturally brings up the issue of exchange
magnetostriction related to a possible lattice tetramerization induced by this Block order, which would reduce the overall energy. Hence, the Block-type order will likely become more stable than the CY-type in some region
by considering the lattice tetramerization. Moreover, according to related DFT calculations and actual
experiments ~\cite{Aoyama:prb19,Zhang:prb18,Dong:PRL14}, the symmetry of the
crystal structure of BaFe$_2$Se$_3$ is reduced due to intra- and inter-ladder lattice distortions.

Thus, next we $include$ the lattice tetramerization in our calculations, where the intra-ladder
Fe-Fe two lattice distances involved are $2.58$ and $2.82$ \AA (for the $Pnma$ phase these numbers are much closer, $2.69$ and $2.72$ \AA). In the lattice tetramerization we include the displacements of Se as well,
due to the exchange striction magnetostriction of iron. All these distortions are confirmed by both theory
and experiment ~\cite{Dong:PRL14,Zhang:prb18,Aoyama:prb19}. By comparing the energies with different magnetic orders, we obtain the new phase diagram shown now in Fig.~\ref{Fig4}(b). The previously remarked small energy difference favoring CY over Block states is now reversed in order, and in the new phase diagram with tetramerization, the CY-type AFM state does not appear in the whole $U-J$ plane.
Instead, the Block-type state becomes more stable because its energy naturally decreases due to the tetramerization. However, it should be noted that the energies of other magnetic orders increase due to the reduced symmetry. In fact, different magnetic orders have different favorable symmetries. Therefore, it is natural that the phase diagram has changed fairly dramatically by considering the lattice tetramerization.

Comparing the different magnetic states using the same lattice arrangement is incomplete because each
particular magnetic order increases its stability -- lowers its energy --  only with the help of a
$particular$ lattice distortion. Thus, the best methodology for further progress would be to fully optimize the crystal for each different magnetic order at various values of $J$ and $U$. However, this is a formidable task.
Given the information we have collected thus far, our best path to arrive to our final
conclusion is to compare the data of the different magnetic states with and without the lattice tetramerization.

The resulting ``combined'' phase diagram is presented in Fig.~\ref{Fig4}(c).
The CY-type state with no lattice distortion remains stable in some portions of the phase diagram, while
the Block state with lattice tetramerization distortion is stable in other regions. The G, CX, and FM states complete the phase diagram. For the widely used ratio $J/U=0.25$, the qualitative tendency with increasing $U$ is first to form a CX-type AFM in a narrow region, followed by a robust Block-type AFM area, and then another robust CY-type AFM region, finally arriving to FM order with further increasing $U$~\cite{FMcontext}.

The proliferation of many competing states at $n=6.0$ as compared with $n=5.5$ probably arises from a combination of correlation effects, increasing Hubbard $U$ and decreasing bandwidth, as well as spin frustrating tendencies between the fully FM state in one
extreme and the purely AFM G-state (in small regions) in the other, as discussed in previous Hartree Fock
calculations \cite{luo:prb13}. However, given the information at hand it can be reasonably assumed that the magnetic state of the $n=6.0$ Te-based iron ladders, if ever prepared experimentally, will not be the CX-type AFM but more likely either the Block- or CY-type arrangements.

\subsection{Projected band structure and density of states }

\begin{figure}[H]
\centering
\includegraphics[width=0.46\textwidth]{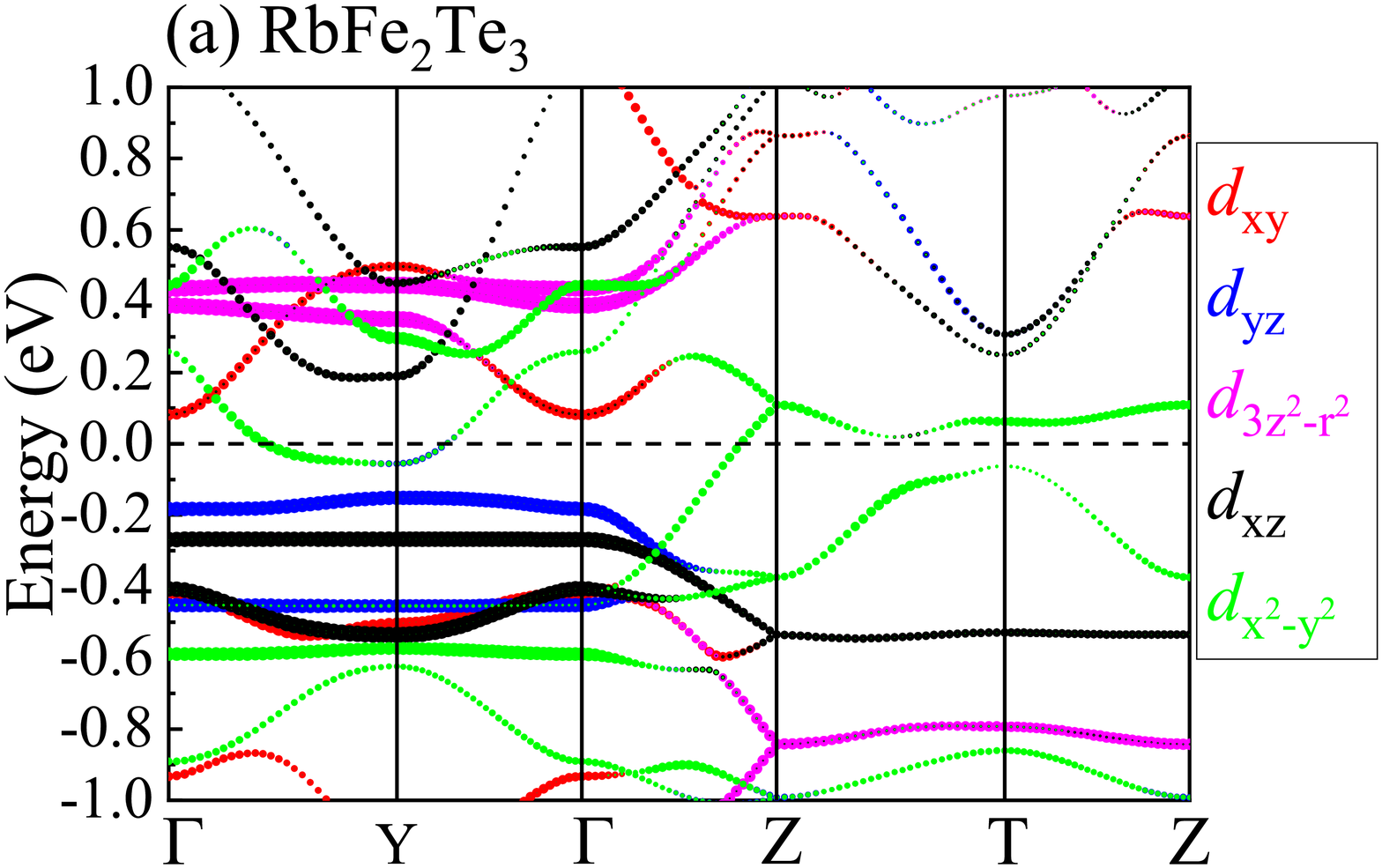}
\includegraphics[width=0.46\textwidth]{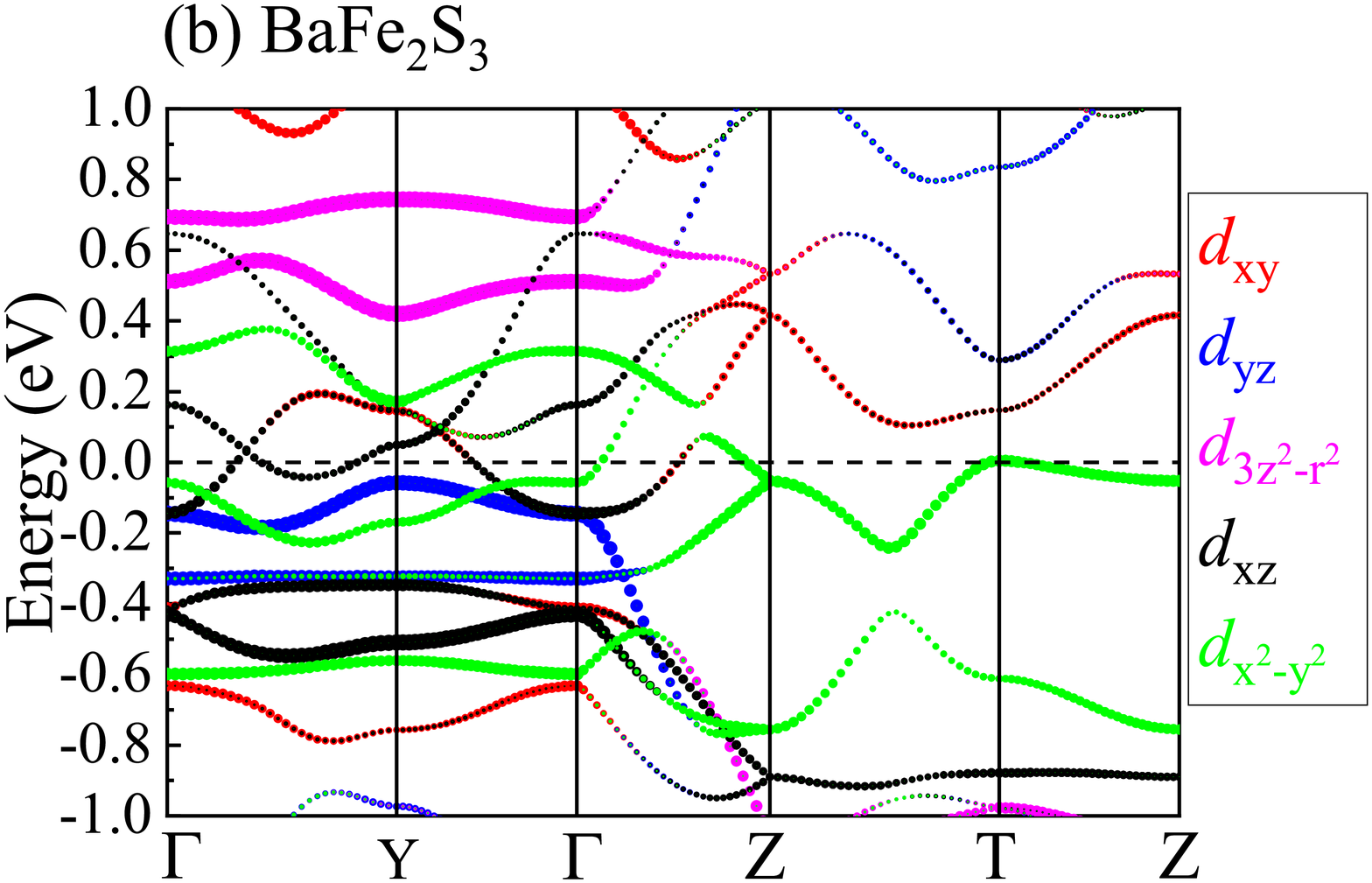}
\includegraphics[width=0.46\textwidth]{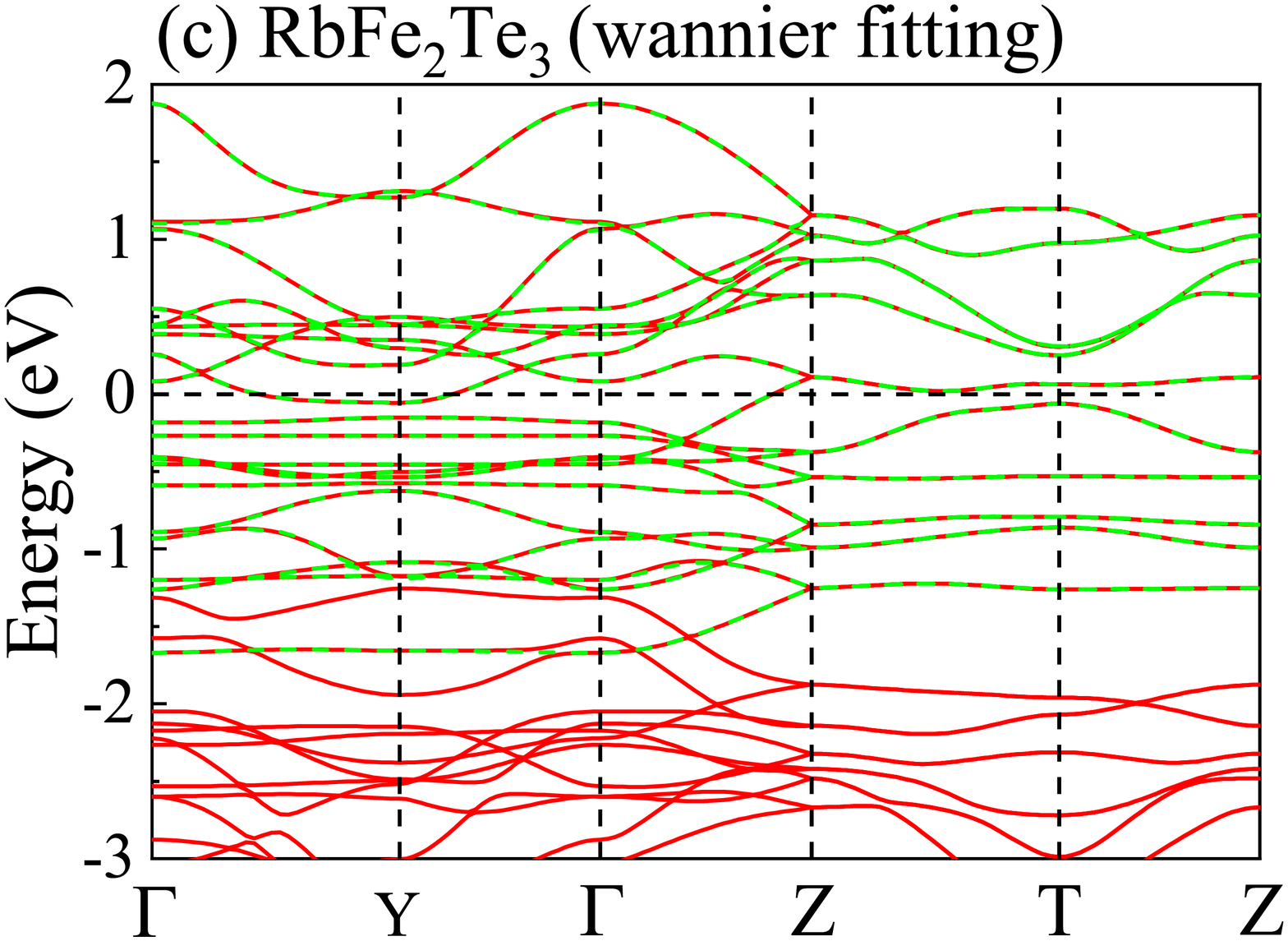}
\caption{(a-b) Projected band structures of RbFe$_2$Te$_3$ (electronic density $n = 5.5$) and BaFe$_2$S$_3$ (electronic density $n = 6.0$) for the non-magnetic (NM) state, respectively. The Fermi level is shown with dashed lines. The weight of each iron orbital is represented by the size of the circle. (c) The original band dispersion is shown by red solid, while the Wannier interpolated band dispersion is shown using green dashed curves for RbFe$_2$Te$_3$.
}
\label{Fig5}
\end{figure}

In Fig.~\ref{Fig5}, we present the ``projected'' band structure of the non-magnetic states restricted only to the five iron $3d$ orbitals corresponding to both RbFe$_2$Te$_3$ and BaFe$_2$S$_3$. It is shown that in general the band structure is more dispersive from $\Gamma$ to Z than along other directions, which is compatible with the presence of quasi-one-dimensional ladders along the $k_z$ axis. We also
observed that the full bandwidth of the five iron $3d$ orbitals
of RbFe$_2$Te$_3$ is smaller than for the case of BaFe$_2$S$_3$, which suggests that the electrons of RbFe$_2$Te$_3$ are more localized than in BaFe$_2$S$_3$. More specifically, the maximally-localized Wannier functions (MLWFs) were employed to fit the five Fe's $3d$ bands by using the WANNIER90
packages~\cite{Mostofi:cpc}. In these Wannier calculations, the bandwidth of the $3d$ orbitals for RbFe$_2$Te$_3$ and BaFe$_2$S$_3$ become approximately $3.56$ eV and $4.06$ eV~\cite{BFScontext}, respectively.

In addition, there are $0.5$ electrons per Fe less in RbFe$_2$Te$_3$ than in BaFe$_2$S$_3$. In RbFe$_2$Te$_3$, the Fermi surface is mainly contributed by the $d_{\rm x^2-y^2}$ orbital while the contribution of the $d_{\rm yz}$ is much reduced, as shown in Fig.~\ref{Fig5}(a) (note that the apparent green color dominance
of the $d_{\rm x^2-y^2}$ bands is misleading: these green bands are actually a mixture of green and blue, the latter arising from $d_{\rm yz}$). The band crossings at the Fermi level along the Y to $\Gamma$
and $\Gamma$ to Z paths have the largest $d_{\rm yz}$ orbital contributions
but always heavily hybridized with the $d_{\rm x^2-y^2}$ orbital. For comparison, in BaFe$_2$S$_3$, the Fermi pockets are mainly contributed by the $d_{\rm x^2-y^2}$, $d_{\rm xy}$, and $d_{\rm xz}$ orbitals
as displayed in Fig.~\ref{Fig5}(b). This clearly suggests that the Fermi pockets of RbFe$_2$Te$_3$
are different from BaFe$_2$S$_3$.

\begin{figure}
\centering
\includegraphics[width=0.48\textwidth]{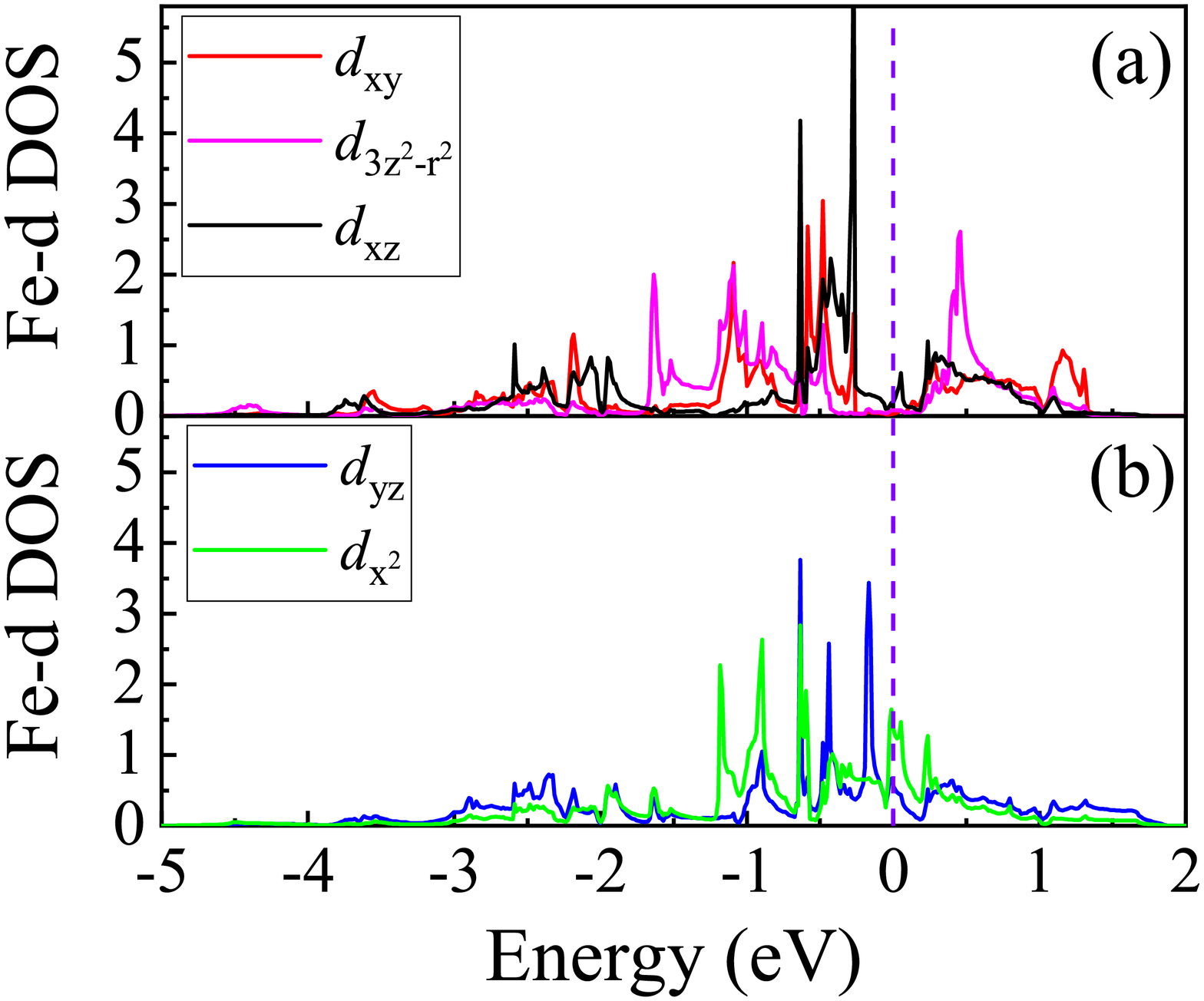}
\caption{The projected density of states of the Fe-$d$ orbitals for the non-magnetic state of RbFe$_2$Te$_3$ (electronic density $n = 5.5$). The five iron $3d$ orbitals are distinguished by different colors.}
\label{Fig6}
\end{figure}

Using the DOS for the Te-ladder (Fig.~\ref{Fig6}), we calculated the relative
proportion of the Fermi surface associated with each of the five iron orbitals:
$64\%$ are contributed by $d_{\rm x^2-y^2}$ and $26\%$ are from $d_{\rm yz}$. For this reason, it seems reasonable to assume that RbFe$_2$Te$_3$ can be described by a two-orbital model or even just one with a combined orbital description ($d_{\rm x^2-y^2}$ hybridized with $d_{\rm yz}$).

Because with increasing pressure the superconducting phase dome of $n=6.0$ BaFe$_2$S$_3$ appears in experiments in the vicinity of the CX-AFM region, the driving force of superconductivity in real systems seems to be the CX spin fluctuations in the nonmagnetic state. According to our previous results for the $n=5.5$ pressured iron ladders ~\cite{Zhang:prb17}, the NM phase can indeed be obtained in theoretical calculations
at high pressure. Due to these similarities, it is reasonable to assume that $n=5.5$ RbFe$_2$Te$_3$ could also
become superconducting at high pressure due to the magnetic similarity with Se- and S-123 ladders,
dominated by the CX state, as it was shown in Fig.~\ref{Fig3}(a).

\section{Discussion}

\begin{figure}
\centering
\includegraphics[width=0.48\textwidth]{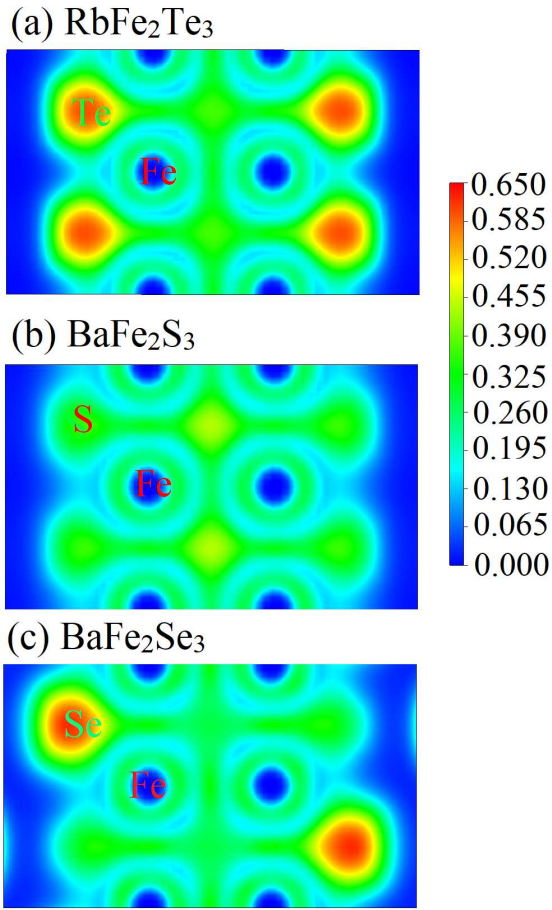}
\caption{The calculated electron localization function (ELF) in the iron ladder plane for (a) $n=5.5$
RbFe$_2$Te$_3$,
(b) $n=6.0$ BaFe$_2$S$_3$, and (c) $n=6.0$ BaFe$_2$Se$_3$, respectively. To better understand the localization
of iron ladders, we set the range of ELF from $0$ to $0.65$.
Generally, ELF=0 indicates no electron localization and ELF=1 indicates full electron localization.}
\label{Fig7}
\end{figure}

Both experimental and first-principles theoretical results revealed a clear tendency for the bandwidths $W$ of
the iron $3d$ orbitals to be enlarged under pressure in BaFe$_2$X$_3$ ~\cite{Arita:prb,Zhang:prb18,Takahashi:Nm}, thus enhancing the itinerant nature of the $3d$ iron electrons. Thus, in this respect pressure reduces the electronic correlation strength given by the ratio $U/W$. To better understand the electronic correlations of RbFe$_2$Te$_3$, we calculated the ``electron localization function'' (ELF) ~\cite{Savin:Angewandte}, quantity widely used within {\it ab initio}
methods to characterize the electron localization. As shown in Figs.~\ref{Fig7}(a) and (b),
the electrons of RbFe$_2$Te$_3$ are more localized than in BaFe$_2$S$_3$, implying that
the electronic correlation of Te-based ladders is stronger. More specifically, electrons in Te are more
localized than in S, and because Te provides the ``bridge'' between irons for the electronic mobility,
then the net effect is that the tunneling amplitude Fe-Te-Fe is reduced as compared with Fe-S-Fe.

Based on the band structure of the NM state, when compared against BaFe$_2$S$_3$ ($\sim4.06$ eV)
the bandwidth of BaFe$_2$Se$_3$ ($\sim3.73$ eV)~\cite{Zhang:prb18} has decreased, which also indicates the electronic correlation
effectively is enhanced. This trend was also observed in our previous theoretical study of the magnetic phase~\cite{Dong:PRL14,Zhang:prb18}.
Hence, it is reasonable to assume the electronic correlation effects for  $n=6.0$ Te ladders -- if they are
ever synthesized -- would be stronger than in BaFe$_2$S$_3$ as well.
Considering also the Block-type AFM order of BaFe$_2$Se$_3$ that is believed to originate in an orbital selective Mott state induced by electronic correlations~\cite{osmp1,osmp2,osmp3}, it is reasonable to conclude that the ground magnetic state of $n=6.0$ Te ladders could display similarly interesting properties.

\section{Conclusion}
In this work, the two-leg iron ladder compound RbFe$_2$Te$_3$, with the iron density $n=5.5$, was
systematically studied using first-principles calculations. The CX-type state was predicted to be the
most likely magnetic ground state. The bandwidths of the iron $3d$ bands in the case of RbFe$_2$Te$_3$ are smaller than in BaFe$_2$S$_3$.

In addition, the phase diagram of ladders at electronic iron density $n=6.0$, corresponding to BaFe$_2$Se$_3$, is found to be much richer than for $n=5.5$. In particular, the 2$\times$2 magnetic Block-type state could be stable at $n=6.0$ according to DFT phase diagrams, particularly due to lattice tetramerization. Also the exotic CY state, with AFM rungs and FM legs, which has $not$ been observed
before neither in experiments nor in calculations, has a large area of stability in the DFT phase diagram at $n=6.0$.

Moreover, considering the predicted dominance of the magnetic CX-state and similarity in electronic structure with other iron ladders, $n=5.5$ RbFe$_2$Te$_3$ may become superconducting under high pressure. Also, according to our ELF analysis, the electrons of Te-123 are more localized than in S, implying that the degree of electronic correlation is effectively enhanced for the Te case, because the Fe-Te-Fe hopping is reduced. This potential relevance of strong correlation
in Te-123 ladders could also induce exotic phenomena, such as the ``orbital selective Mott physics'' recently
discussed using multiorbital Hubbard models~\cite{osmp1,osmp2,osmp3}. Our overarching conclusion is that experimental studies of iron ladder tellurides are worth pursuing, because using Te could lead to interesting results, such as exotic magnetic states and superconductivity under high pressure.

\section{Acknowledgments}

E.D. and A.M. are supported by the U.S. Department of Energy (DOE), Office of Science, Basic Energy Sciences (BES), Materials Sciences and Engineering Division. S.D., Y.Z., and L.F.L. were supported by the National Natural Science Foundation of China (Grant Nos. 11834002 and 11674055). L.F.L. and Y.Z. were supported by the China Scholarship Council. Y.Z. was also supported by the Scientific Research Foundation of Graduate School of Southeast University.  Most calculations were carried out at the Advanced Computing Facility (ACF) of the University of Tennessee Knoxville (UTK).


\begin{references}
\bibitem{Jcontext} Here, we qualitatively obtained the Heisenberg exchange couplings $J$ by comparing the energies of several collinear magnetic orders. Choosing different sets of various magnetic states will result in different values of $J$. However, it will not change the qualitative tendencies.
\bibitem{note-added} In model calculations, there is a third
coupling called $U'$ representing the repulsion
between electrons at different orbitals in the same atom. Rotational invariance imposes a constraint $U$=$U'$+2$J_H$
between the on-site same-orbital Hubbard $U$, the model Hund coupling $J_H$, and $U'$~\cite{Patel:prb16}.
This restricts the range
of $J_H/U$ to be less than 1/3, otherwise if larger than 1/3 then $U'<J_H$ considered unphysical. Here, we include $J/U>1/3$ for completeness, but none of our physical conclusions change if only $J/U<1/3$ is studied.
\bibitem{BFScontext} This value is obtained from our previous calculations of BaFe$_2$S$_3$, which is consistent with other publication
as found in~\cite{Arita:prb}.
\end{references}
\end{document}